\documentclass{PoS}

\usepackage{amsmath}
\usepackage{amsfonts}
\usepackage{mathrsfs}
\usepackage{amssymb}
\usepackage{psfrag}

\newcommand{\Z}{\mathbb{Z}}

\newcommand{\bra}{\langle}
\newcommand{\ket}{\rangle}
\newcommand{\avg}[1]{\langle \hspace{0.2em} #1 \hspace{0.2em} \rangle}

\newcommand{\eq}{\begin{equation}}
\newcommand{\en}{\end{equation}}
\newcommand{\bea}{\begin{eqnarray}}
\newcommand{\ea}{\end{eqnarray}}
\newcommand{\arccosh}{{\rm Arccosh}}
\newcommand{\link}[1]{< \hspace{-0.18em} #1 \hspace{-0.18em} >}

\title{Broadening of the confining flux tube joining higher-representation sources}

\ShortTitle{Broadening of the confining flux tube joining higher-representation sources}

\author{Pietro Giudice, Ferdinando Gliozzi, \speaker{Stefano Lottini}\\
        Universit\`a di Torino and INFN, sez. di Torino, Italy\\
        E-mail: \email{giudice},
        \email{gliozzi},
        \email{lottini}\:\email{@to.infn.it}}

\abstract{In the context of confining gauge theories we study the flux tube generated by a pair of static sources belonging to higher rank representations of the gauge group. Using a simple geometric approach based on minimal surfaces describing the world-sheet of the underlying k-string, we infer logarithmic broadening of the flux tube as a function of source separation. It turns out that the coefficient of the logarithm does not depend on the specific representation of the source nor on its N-ality and is universal. Numerical results on a $\Z_4$ gauge theory in three dimensions with static sources carrying two units of elementary flux compare very favourably with these predictions.}

\FullConference{XXIVth International Symposium on Lattice Field Theory\\
                July 23-28, 2006\\
                Tucson, Arizona, USA}

\begin{document}

It is widely believed that a confining gauge theory near the continuum limit 
is in  the rough phase. This means, as explained long ago by 
L\"uscher, M\"unster and Weisz \cite{lmw}, that the colour flux tube  
broadens logarithmically with source separation. This phenomenon can be seen 
 as a consequence of the underlying string description of the infrared 
properties of the flux tube. The detailed form of the string action is not 
know, however its IR limit is expected to be a two-dimensional massless 
Gaussian model (free string) made with $d-2$ bosonic fields 
$\varphi_\perp(\xi_1,\xi_2)$  describing the transverse fluctuations 
of the world-sheet having a Wilson loop $W(C)$ as boundary.  These quantum 
fluctuations give rise to two universal effects: one is a 
contribution to the vacuum expectation value of the Wilson loop $\bra W(C)\ket$
which depends only on the shape of $C$. It leads in particular to the 
L\"uscher term
in the static potential. These kinds of shape corrections  have been 
observed in many gauge systems in three  and four dimensions for various 
gauge groups. 

The other universal effect is just the logarithmic growth of the mean 
square width of the colour flux tube. Observing this broadening is very 
challenging from a computational point of view. As a matter of fact, 
an uncontroversial observation of it has been 
made only in 3D $\Z_2$ gauge model, thanks to the efficiency of the 
Monte Carlo algorithms for its dual, the Ising model \cite{cgmv}.
 
The question naturally arises whether such a broadening  is present also in
the k-strings i.e. the effective strings describing the confining force between
static sources carrying the quantum numbers of k copies of the fundamental 
representation. These strings can be considered as bound states of 
k fundamental strings. It is not clear whether in the IR 
limit the only relevant degrees of freedom are the transverse displacements
of a single string or whether we have to add some new field describing the 
possible splitting of the k-string into its constituent strings. Therefore
 we are not entitled to apply blindly the IR gaussian description of the 
fundamental string to the k-string issue.

\section{K-strings and minimal surfaces}

To inspect the width of the flux tube between two static sources in the fundamental representation one starts from the quantity
\eq
	P_f(h) = \frac{ \avg{W_f(C)W_f(c)} - \avg{W_f(C)}\avg{W_f(c)} }{\avg{W_f(C)}}
\en
where $W_f(C)$ and $W_f(c)$ are Wilson operators for the loops $C$ and $c$, which are coaxial circles of radii $R$ and $r < R$ lying on planes separated by a length $h$. The correlator above defined can be seen as the transversal density of the colour flux generated by the source $C$ in the fundamental representation and measured with the probe $c$. To study the flux generated by sources in a generic representation $\mathscr{R}$, we will take $W_{\mathscr{R}}(C)$ instead of $W_f(C)$, but keep the probe always in the fundamental.

The mean squared width of the flux tube is defined by
\eq
	w^2_{\mathscr{R}}=\frac{\int h^2 P_{\mathscr{R}}(h)\mathrm{d}h}{\int P_{\mathscr{R}}(h) \mathrm{d}h}
	\label{eq:defw2}
\en

The insight of \cite{lmw} was to observe that the quantity $P_f(h)$ can be described in the effective string picture by a typical Plateau problem of minimal surfaces, which can then be evaluated in the infrared limit by a saddle-point approximation. More specifically, one can write
\eq
	\avg{W_f(C)W_f(c)} - \avg{W_f(C)}\avg{W_f(c)} \propto \exp(-\sigma A(R,r,h))
\en
where $\sigma$ is the string tension and $A(R,r,h)$ denotes the area of the connected minimal surface with $C$ and $c$ as boundaries. Here it is assumed that the effective string is described by the Nambu-Goto action; it is reasonable to expect that in the infrared limit $R\to \infty$ with fixed $r$ and $h$ the result should not depend on this specific assumption. Note however that the size $r$ of the probe cannot be too small, otherwise the string picture would not be valid.

The surface with minimal area connecting the two circles is a surface of revolution about the symmetry axis (here chosen to coincide with the $\boldsymbol{x}$ axis). If we denote by $y(x)$ the $y>0$ intersection of the surface with the $(x,y)$ plane, the area is given by
\eq
	A = 2\pi \int y \sqrt{\dot{y}^2+1} \mathrm{d} x
\en
\vspace{-1.7cm}
\begin{center}
\raisebox{1cm}{\includegraphics[width=3cm]{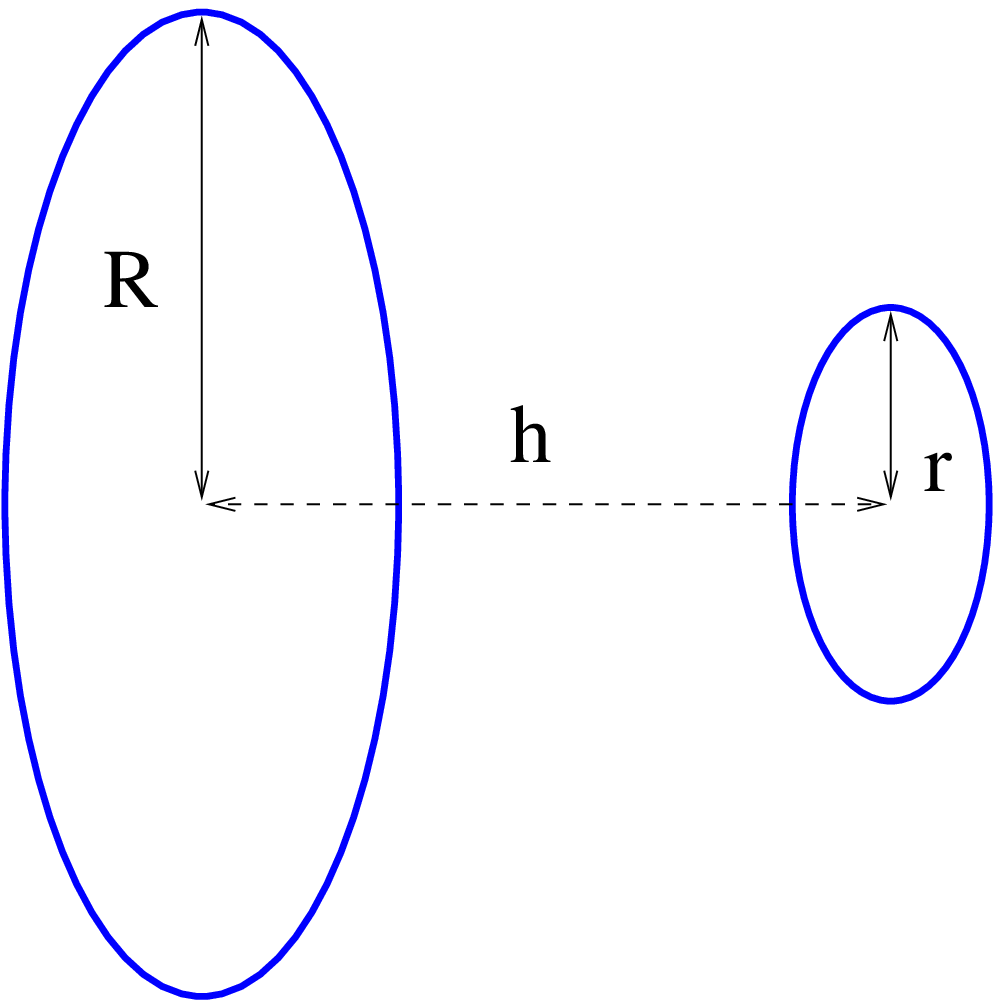}}\quad\quad\quad\includegraphics[width=7cm]{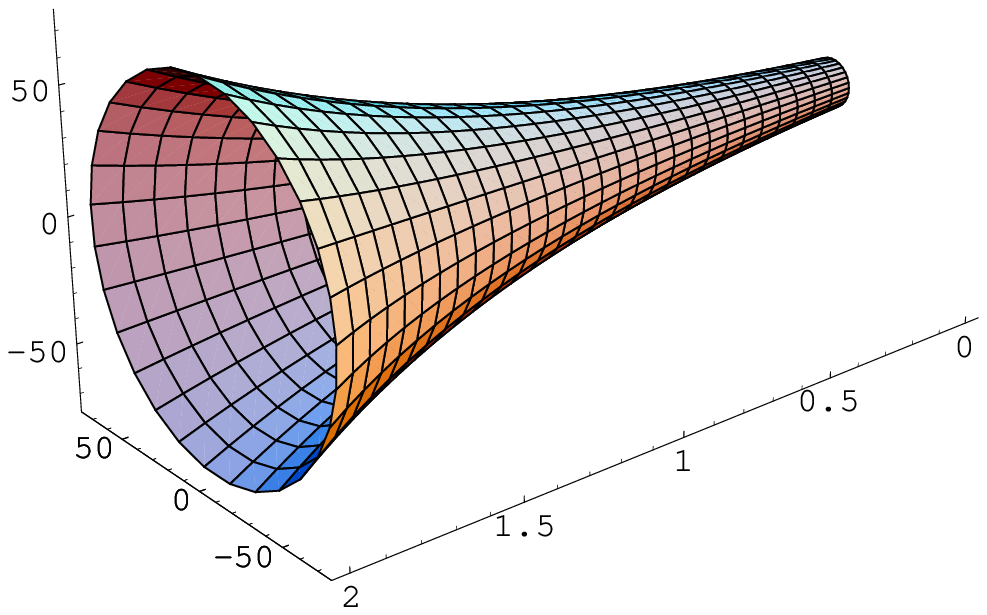}
\end{center}

By imposing the extremal condition $\frac{\delta A}{\delta y(x)}=0$ one finds the general solution:
\eq
	1+\dot{y}^2 = y \ddot{y} \quad \Rightarrow \quad y(x) = \frac{1}{\omega} \cosh\omega(x-x_0)
\en

Applying the boundary conditions $R = \frac{1}{\omega}\cosh\omega x_0$, $r=\frac{1}{\omega}\cosh\omega(h-x_0)$, it is possible to express the minimal area as:
\eq
	A = \pi\Big(\frac{h}{\omega}+R^2\sqrt{1-1/\omega^2R^2}-r^2\sqrt{1-1/\omega^2r^2}\Big)
\en
From the boundary conditions we can express the distance $h$ as a function of $\omega$ as
\eq
	h(\omega)=\big[\arccosh(R\omega)-\arccosh(r\omega)\big]/\omega
	\label{eq:accadiomega}
\en
which is a decreasing, hence invertible, function: we can use $\omega$ as an integration variable for actually computing $w^2$. Notice that since the hyperbolic cosine is always $\geq 1$ we get a minimal value for $\omega$, that is a maximal allowed value for $h$:
\eq
	\omega_{min} = \frac{1}{r} \quad \Rightarrow \quad h_{max} = h(1/r)
\en
Now we can write explicitly, for the mean squared width,
\eq
	\avg{w^2}=\frac{\int_{\frac{1}{r}}^\infty h(\omega)^2\,exp[-\sigma A(R,r,h(\omega))]
	\vert h'\vert\,d\omega}{
	\int_{\frac{1}{r}}^\infty \,exp[-\sigma A(R,r,h(\omega))]
	\vert h'\vert\,d\omega}
\en

This quantity approaches a logarithmic curve for large $R$. Indeed, this can be seen by expanding $\arccosh(\omega r) \sim 2 \log(\omega r)$ for $\omega r \gg 1$, and inserting it into (\ref{eq:accadiomega}), which yelds the approximate solution found in \cite{lmw}:
\eq
	\omega \sim \frac{1}{h}\log(R/r)
\en
The condition $\omega r \gg 1$ becomes $\log (R/r) \gg h/r$, always fulfilled in the large $R$ limit (moreover, $r$ cannot be too small at fixed $R$ and $h$). In this limit, a Gaussian distribution is found for the transversal density, whose width grows logarithmically with $R$:
\eq
	P^{(R)}(h) \propto exp\Big[-\frac{\sigma \pi h^2}{\log(R/r)}\Big] \quad \Rightarrow \quad
               \sigma w^2 = \frac{1}{2 \pi} \log (R/r)
\en

The result found here, in full agreement with the one from the conformal field theory approach to the problem, can be generalised to Wilson loop $W_\mathscr{R}(C)$ in any representation. The worldsheet of the k-string can be seen as some bound state of $k$ 1-string worldsheets. In this case a local minimum solution exists, in which one of these fundamental sheets gives rise to the catenoid with the probe $W_f(c)$, while the other $k-1$ just lie flat on the loop plane. We then have immediately that the resulting width does not change at all from the $k=1$ case, since
\eq
	P^{(R)}_{(k)}(h) = P^{(R)}_{(1)}(h) \cdot \exp[\pi R^2 (\sigma_k - \sigma_{k-1} - \sigma)]
\en

\section{$\Z_4$ gauge theory and duality to the Ashkin-Teller model}

The laboratory where we study the physical properties of the 2-string is a $\Z_4$ gauge model in three dimensions, defined by the standard plaquette action on a cubic lattice: with $c_1$ and $c_2$ coupling constants, and with link field $U_\ell \in \Z_4 = {+1,+i,-1,-i}$,
\eq
	S = \sum_P \Big[ c_1 \mathfrak{Re}(U_P) + c_2 (U_P)^2 \Big] \: .
\en
With $\Z_4$ as gauge group, there exist two k-strings in the system: the fundamental string and the 2-string (related to the double-fundamental representation $f\otimes f$). Moreover, since there are no more representations than one in the same N-ality class, the system does not exhibit any meta-stable string that could spoil the results at finite $R$.

It is well known that this model, as any three-dimensional abelian gauge model, admits a spin model as its dual (see for instance \cite{id}) and any physical property of the gauge system can be translated into a corresponding property of its spin dual. From a computational point of view it is of course much more convenient to work directly on the spin model where powerful non-local cluster algorithm can be applied.

In our case the dual is a spin model with global $\Z_4$ symmetry which can be written as a symmetric Ashkin-Teller (AT) model \cite{at}, i.~e.~two coupled, ferromagnetic, Ising models defined by the two-parameter action
	\eq
	S_{AT} = \sum_{<xy>} \Big[ \beta (\sigma_x \sigma_y + \tau_x \tau_y) 
                 + \alpha (\sigma_x \sigma_y \tau_x \tau_y) \Big]
	\en
where $\sigma_x$ and $\tau_x$ are the Ising variables ($\sigma,\tau=\pm 1$) associated to the site $x$ and the sum is over all the links $\link{xy}$ of the dual cubic lattice. The phase diagram of this model has been studied long ago \cite{dbgk,az}. The global $\Z_4$ symmetry of the action is generated by the transformation \makebox{$\sigma \to - \tau \: ; \: \tau \to \sigma$}; the model has also an independent $\Z_2$ symmetry generated by \makebox{$\sigma\leftrightarrow\tau$}, related to the charge conjugation of the corresponding dual model.

Usually the $\Z_4$ symmetry is realised by means of fields with values among the fourth roots of the identity ($\xi_j = \pm i, \pm 1$): the double $\Z_2$ formulation corresponds to the standard one via the change of fields \makebox{$\Psi_x=e^{-i\frac{\pi}{4}}(\sigma_x+i \tau_x)$}, which preserves the above symmetries as $\Psi \leftrightarrow \Psi^\star$ and the multiplication by $i$. We think it is interesting (and more purposeful for our work) to prefer the formulation with two $\Z_2$ coupled fields: indeed, using character expansion for the Boltzmann weights and resumming in terms of newly introduced (dual) link variables, it can be shown that the AT model is dually equivalent to \emph{two coupled} $\Z_2$ gauge models with couplings $a$, $b$ ($U_P$ and $V_P$ are the two $\Z_2$-like plaquette operators)
\eq
	Z_{AT}(\beta,\alpha) \propto \sum_{ \{ U_\ell = \pm 1 , V_\ell = \pm 1 \} } \prod_P e^{b(U_P+V_P)+a U_P V_P}
\en
The duality is implemented by the transformation $\mathscr D : (\alpha, \beta) \to (a,b)$, which satisfies $\mathscr{D}^2 = 1$:
	\eq
	(a,b)= \Bigg( \frac{1}{4} \ln \Big[ 
	\frac{(\coth\beta+\tanh\beta\tanh\alpha)(\coth\beta+\tanh\beta\coth\alpha)}
        {2+\tanh\alpha+\coth\alpha} \Big] , \frac{1}{4} \ln \Big[
	\frac{1+\tanh^2\beta\tanh\alpha}{\tanh^2\beta+\tanh\alpha} \Big] \Bigg) \: .
	\en

\subsection{Dual of a Wilson loop}

The duality transformation maps any physical observable of the gauge theory into a corresponding observable of the spin model. In particular it is well known that the Wilson loops are related to suitable \emph{twists} of the couplings of the spin model. More specifically, let us consider a $\Z_N$ spin model in 3D. Let $\xi=e^{i 2 \pi / N}$ be the generator of this group. A $k$-twist of the link $\link{xy}$ in the spin action is defined by the substitution $\beta \to \xi^k \beta$ only in the selected link. Denoting with $Z_{<xy>,k}$ the spin partition function modified in this way, one can easily prove the identity
\eq
	\avg{U_P^{(k)}}_{gauge} = \frac{Z_{<xy>,k}}{Z}
\en
where $P$ is the plaquette dual to $\link{xy}$ and $U_P^{(k)}$ is the plaquette variable in the irreducible representation of $\Z_N$ characterised by the integer $k = 1,2,\ldots,N$. As a simple check, by twisting the six links dual to an elementary cube, one obtains the same partition function due to the invariance of the measure; on the gauge side, this is equivalent to measuring the product of the six plaquettes of the cube, which is always 1.

In the AT model, the twist on a link can be performed regarding both $\Z_2$ components or, say, the $\sigma$ field. What results from the above recipe is that twisting only one component corresponds to operators in the fundamental representation, while twisting both is involved in the double-fundamental. Thus, in the AT action, we have (specifically for the frustrated link dual to the plaquette of interest)
\mbox{$[\beta (\sigma_x \sigma_y + \tau_x \tau_y)+\alpha (\sigma_x \sigma_y \tau_x \tau_y)]
\to[\beta (-\sigma_x \sigma_y + \tau_x \tau_y)-\alpha (\sigma_x \sigma_y \tau_x \tau_y)]$} for the fundamental representation, and 
\mbox{$[\beta (\sigma_x \sigma_y + \tau_x \tau_y)+\alpha (\sigma_x \sigma_y \tau_x \tau_y)]
\to [-\beta (\sigma_x \sigma_y + \tau_x \tau_y)+\alpha (\sigma_x \sigma_y \tau_x \tau_y)]$} for the double-fundamental. Combining together a suitable set of plaquettes we may build up any Wilson loop or Polyakov-Polyakov correlator with $k=1$ or $k=2$ and their map into the spin model. Also, these are changes in the action that can be embedded in the simulation algorithm.

The actual measure of a plaquette operator $V_P$ translates in magnetic terms as
\bea
	\avg{V^{(k=1)}_P}_{gauge} & = & \avg{e^{-2(\beta+\alpha\tau_x\tau_y)\sigma_x\sigma_y}}_{AT} \\
	\avg{V^{(k=2)}_P}_{gauge} & = & \avg{e^{-2\beta(\sigma_x\sigma_y+\tau_x\tau_y)}}_{AT}
\ea
In practice, to measure a plaquette in the fundamental representation, it is a good approximation, at small $\alpha$, use the formula (where $\epsilon$ accounts for possible frustrations on the $\sigma$ component of the link):
\eq
	\avg{U_p^{f}}_{gauge} \stackrel{\sim}{\propto} \avg{\epsilon \sigma_x \sigma_y}_{AT} + \mathrm{constant}
\en

\section{Monte Carlo analysis}

We performed a Monte Carlo analysis on the AT model, for both the fundamental and the double-fundamental strings, at the (confining) coupling $(\alpha,\beta)=(0.0070,0.1975)$ (for which we have measured the string tensions $a^2\sigma=0.01560(1)$ and $a^2\sigma_2= 0.0210(6)$).

On the lattice, the quantity $P_\mathscr{R}^{(R)}(h)$ is measured by placing a square $R$ by $R$ loop $W(R)$ on a plane in the desired representation and taking the probe as a plaquette operator, parallel to the loop and lying on its axis at a distance $h$. The actual measurement takes into average also the four planar neighbours of the plaquette in the central position, in order to enhance the signal; this operation does not spoil the results since we dealt with large values of $R$.

The large loop is embedded into the update procedure as a set of frustrated links; the algorithm has only to measure the average value of the probe plaquette in the fundamental representation, since $P_\mathscr{R}^{(R)}(h) \propto \avg{U_p^{f}}^{W_\mathscr{R}} + \mathrm{constant}$.

The algorithm used for the analysis uses a cluster update method \cite{wd} basically similar to the standard Fortuin-Kasteleyn cluster technique: each update step is composed by an update of the $\sigma$ variables using the current values of the $\tau$ as a background (thus locally changing the coupling from $\beta$ to $\beta \pm \alpha$ according to the value of $\tau_x \tau_y$ on the link $\link{xy}$), followed by an update of the $\tau$'s using the $\sigma$ values as background. We reached a performance of about 0.35 seconds per update/measurement step (as on a single Intel$^{\mathrm{\scriptsize{\textregistered}}}$ Xeon 3.2 GHz 64-bit processor).

We used a cubic $L^3$ lattice with side $L=80$ (we found there are no finite size deviations up to $R \simeq L/2$) and measured $N$ times the plaquette operator as discussed. Configuration results have been then packed in groups of 25 in a binning fashion to estimate variances. We performed the measurements on loop sides $R=11,13,\ldots,41$ with a statistics of 220875 measures (independently for each $R$) for the fundamental representation and 470275 for the $k=2$.

\subsection{Results}

We found that the measured values of $P_\mathscr{R}^{(R)}(h)$ do not fit too well to a normal distribution as approximately expected and lead to bad width estimates, so we used the numerically integrated quantities instead, as in Eq.~(\ref{eq:defw2}). We performed the integration by carefully choosing a cutoff value $h_{max}$: since the (indipendently measured) background value must be subtracted from the transverse flux density functions, the results are very sensitive to the choice of a cutoff. We took $h_{max}=19$.
\vspace{-1cm}
\begin{center}
\includegraphics[width=8cm,angle=270]{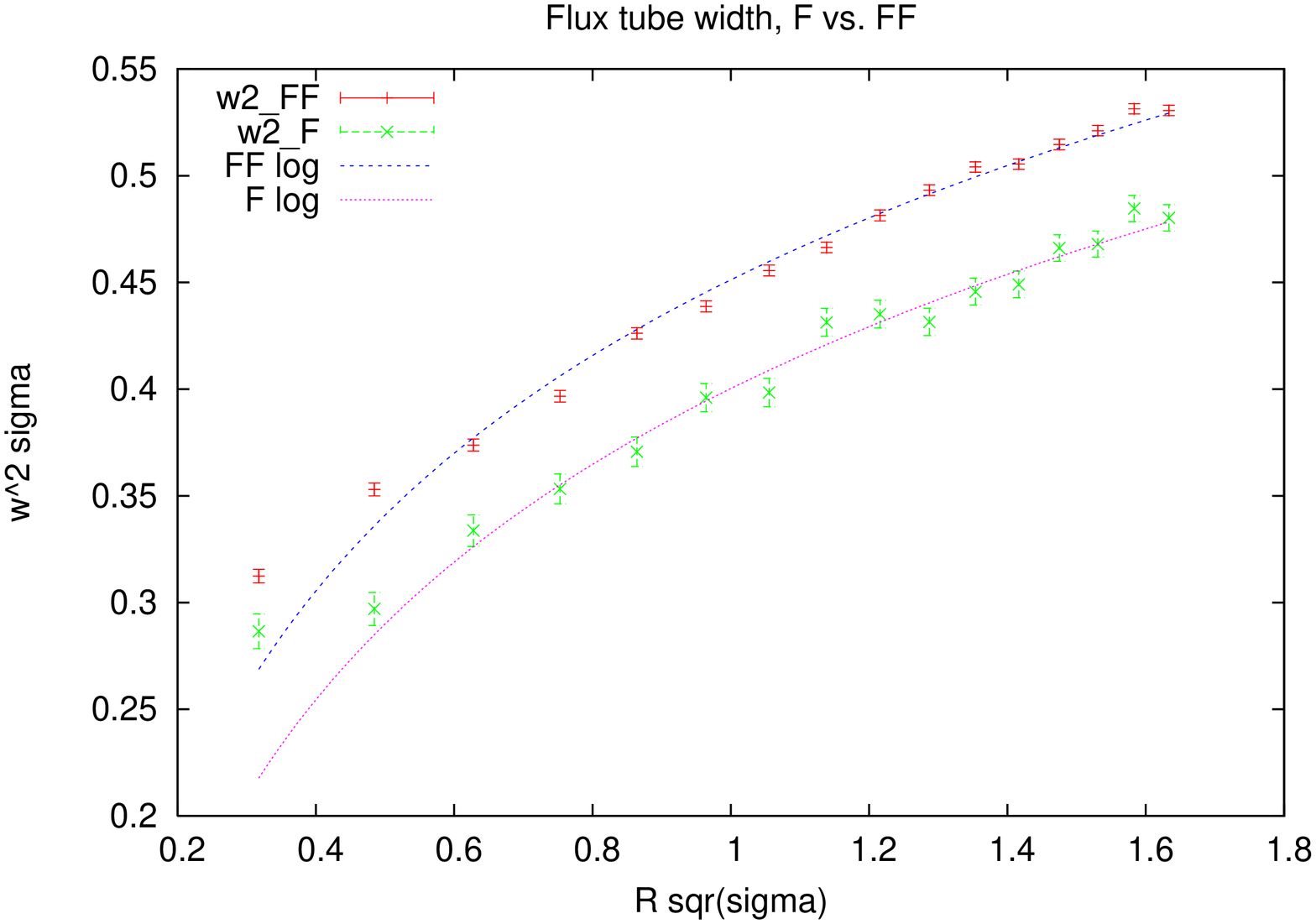}
\end{center}

By fitting the functions $w^2_{k=1,2}(R)$, for $R\geq R_{min}$ with $R_{min}$ an appropriate distance cutting off non-IR contributions, to the functional form
\eq
	w^2(R) = \frac{1}{2\pi}\log R + c
\en
we found that the fundamental string width, as well as the 2-string, show the expected logarithmic growth with the appropriate universal multiplicative factor (reduced $\chi^2$ were, respectively, 1.22 and 2.68), but the value of $c$ differs measurably in the two cases: $c_1=0.4002(20)$, $c_2=0.4512(11)$; this discrepancy is probably due to some interaction between the fundamental worldsheets in the $k=2$ case.

Our work reinforces the numerical evidences of the broadening of the flux tube width \cite{cgmv}, extending them with high precision to the case of $Z_4$. The main result is that this is valid also for non-fundamental strings: the assumption of free massless string infrared limit appears confirmed.

\end{document}